# PAMBE growth of GaN nanowires on metallic ZrN buffers - a critical impact of ZrN layer thickness on the substrate surface temperature

K. Olszewski[1], Z.R. Zytkiewicz[1,*], A. Wierzbicka[1], M. Guziewicz[2], and M. Sobanska[1,*]

[1] Institute of Physics, Polish Academy of Sciences, Al. Lotnikow 32/46, 02-668 Warsaw, Poland

[2] Łukasiewicz Research Network - Institute of Photonics and Microelectronics, Al. Lotnikow 32/46, 02-668 Warsaw, Poland

[*] corresponding authors: sobanska@ifpan.edu.pl; zytkie@ifpan.edu.pl

## Abstract

An impact of thin metallic ZrN layers on Si and sapphire wafers on substrate temperature during MBE growth of GaN nanowires is studied. Using nucleation kinetics of GaN as a sensitive probe we show that a thin ZrN layer strongly increases the substrate temperature, which significantly affects the dimensions and density of the nanowires. To quantify the effect we developed a technique of optical pyrometer calibration that allows reliable determination of emissivity, and thus precise measurement of temperature of substrates with unknown optical parameters, such as ZrN buffers of various thicknesses. Our results show that emissivity of ZrN-coated Si and sapphire wafers differs significantly from the bulk ZrN and increases drastically for films thinner than ~100 nm. Simple calculations indicate that ignoring the influence of the thin film may lead to huge errors in temperature readings and consequently to losing the growth control. Then, we show that we can compensate for the impact of ZrN buffer on substrate temperature and grow identical nanowire arrays on Si substrates with and without ZrN layers. Finally, having identical arrays of GaN nanowires we used X-ray diffraction to compare nanowire arrangements on Si and ZrN/Si substrates with a thin SiN nucleation layer.

## Introduction

III-nitride nanowires (NWs) provide an attractive alternative for planar structures in micro- and optoelectronic devices. Due to a small footprint in contact with a substrate, nanowires are free of misfit dislocations even when grown on highly lattice-mismatched substrates [1]. As a result, nanowires are of superior structural quality which is not achievable in comparable planar heterostructures [2–4]. The process of self-assembled growth of GaN NWs by molecular beam epitaxy (MBE) has been thoroughly investigated, especially for amorphous substrates such as nitridated Si [1,4–9], $SiO_x$ [10,11] and $Al_xO_y$ [12–14]. A specifically attractive benefit offered by the self-assembled growth of GaN NWs on such substrates is that the epitaxial constraints are very weak and the NWs grow perpendicular to the substrate surface [9]. Consequently, ensembles of well-aligned vertical NWs of equal height are formed, which is crucial for the subsequent processing and fabrication of NW-based light emitting devices such as LEDs. Unfortunately, $SiO_2$, $Al_xO_y$ and $SiN_x$ buffer layers form non-ohmic electrical contacts to GaN. This hinders carrier transport and heat dissipation at the NW/substrate interface [15]. Furthermore, due to the optical transparency of $SiO_2$, $Al_xO_y$ and $SiN_x$ buffers, a large part of light generated in NW-based LEDs is lost by its absorption in a Si carrier wafer. Thus, various alternative substrates for GaN NW-based devices are still tested.

There is an increasing interest in growing GaN NWs on metal substrates [16–24]. Such substrates exhibit excellent electrical and thermal conductivities as well as a high optical reflectance. However, reports on NW growth on metals such as Mo, Ti, Ta, and W [16,18,25] show the instability of the metal layers during MBE growth [18]. Therefore, more stable metallic nitride substrates, as ZrN, TiN, etc., attract much attention. In particular, the high stability of ZrN was confirmed by our recent study, in which we demonstrated arrays of vertically oriented GaN NWs grown by plasma-assisted MBE (PAMBE) on ZrN layers sputtered on Si carrier wafers [26]. Moreover, low resistivity ohmic electrical contact of GaN NWs to ZrN has been observed [27]. Thus, in electrically driven devices the nucleation ZrN layer can effectively play a role of low resistive electrical contact to bottom part of the nanowire ensemble. Such unique properties of metallic nitrides should lead to a significant enhancement of the efficiency of NW-based optoelectronic devices built on such substrates.

Fabrication of device relevant structures requires a precise control of nucleation and growth of GaN nanowires. In particular, substrate surface temperature must be accurately adjusted as the NW incubation time increases exponentially with temperature [13,14,28,29], so the growth temperature critically affects nanowire nucleation and growth kinetics.

The most common method for determining the surface temperature of a substrate during MBE growth is optical pyrometry [30,31]. By measuring the intensity of infrared radiation emitted by a heated object, a pyrometer allows for a noncontact evaluation of the substrate temperature. This approach is nondestructive and allows for continuous temperature measurement of a rotating substrate, making it perfectly suitable for MBE technique. However the greatest challenge in utilizing this method is the correct correlation between the real substrate temperature and the pyrometer reading [32–34]. The intensity of infrared radiation emitted by a heated object directly depends on its emissivity which is an inherent value that varies for different materials. Moreover, secondary factors such as the optical transmission of the pyrometer viewport in MBE system and its time variation caused by a progressing coating of the viewport also impact pyrometer readings [35]. Therefore precise pyrometer calibration protocols must be developed and periodically applied. This is especially important if the growth proceeds on thin metallic buffer layer since emissivity of such buffers is unknown and, as we show in this work, strongly depends on the layer thickness. Using metal bulk emissivity values and disregarding the influence of the thin film may lead to significant errors in the substrate temperature measurement and consequently to the loss of control of the growth process.

In this work we show experimental evidences that presence of thin ZrN layer significantly affects substrate temperature during MBE growth. The effect is demonstrated by comparing the self-assembled growth of GaN nanowires by PAMBE on a dedicated Si substrate partly covered by a thin ZrN buffer. The incubation time of GaN nanowires is strongly temperature-dependent, so the kinetics of their nucleation is a sensitive probe of even small changes in the surface temperature of the substrate. Using such a probe we show that the local presence of 40 nm thick ZrN film on Si carrier wafer increases the growth temperature by as much as 17ºC, which significantly affects the dimensions and density of the nanowire array for exactly the same heater power.

To quantify the effect and regain control over the growth we used an optical pyrometry-based temperature calibration technique via melting of aluminum wires attached to the substrate that allows reliable determination of emissivity, and thus precise measurement of surface temperature of custom substrates coated with exotic materials with unknown optical parameters, such as ZrN buffer layers of various thicknesses deposited on either silicon or sapphire wafers. Then, we show that by using our temperature calibration procedure we can compensate for the impact of ZrN buffer on substrate temperature and grow identical GaN nanowire arrays on two Si substrates with and without ZrN layers. This finding provides a direct evidence that we can accurately designate the desired temperature in separate PAMBE growth processes on such different substrates and confirms again the usefulness of our approach. On a general note, we emphasize that our temperature calibration technique is universal, simple, and MBE compatible. As such it should also be applicable to entirely different material systems for both types of grown nanocrystals as well as host wafers and deposited thin buffer layers. Finally, having two identical arrays of GaN NWs we used x-ray diffraction to compare arrangements of the NWs grown on Si and ZrN/Si substrates with a thin SiN nucleation layer.

**Experimental**

ZrN buffer layers were deposited on either Si(111) or c-plane $Al_2O_3$ carrier wafers using DC sputtering from the ZrN target. Sputtering was performed at room temperature under Ar pressure of 1.4 x $10^{-3}$ mbar and deposition rate of 0.13 nm/s. Additionally, the back surfaces of $Al_2O_3$ substrates were covered with a 250 nm thick Mo layer to increase heat absorption from the heater. All the substrates were transferred in air to the RIBER Compact 21 PAMBE system and subsequently heated to 150°C for 1 h in the loading chamber to remove any volatile contaminants. Substrates dedicated for the growth of nanowires were additionally heated to 500°C for 5 h in the preparation chamber. GaN nanowires were grown for 120 min with constant N and Ga fluxes of $\Phi_N$ = 15 nm/min and $\Phi_{Ga}$ = 8 nm/min, respectively. Cross-sectional scanning electron microscopy of thick GaN(0001) films grown under slightly N- and Ga-rich conditions at low temperatures (680°C) was used to calibrate the Ga and N fluxes in the GaN growth rate equivalent units of nm/min. The stability of the N flux was controlled during growth using an optical sensor of plasma light emission attached to the plasma cell [36]. The substrate temperature was measured with a Raytek Marathon MM2ML single colour infrared pyrometer operating at the wavelength of 1.6 µm and focused at the substrate surface to the spot size of ~3 mm. After growth, the height and density of the NW arrays were examined with a Zeiss Auriga scanning electron microscope (SEM). X-ray diffraction (XRD) was used to determine the crystallographic arrangement of the GaN NWs relative to the substrate. X-ray measurements were performed with a Panalytical X'Pert Pro MRD X-ray diffractometer equipped with a hybrid two bounce Ge (220) monochromator with an X-ray mirror, a threefold Ge (220) analyser and a Pixel detector.

**Results and discussion**

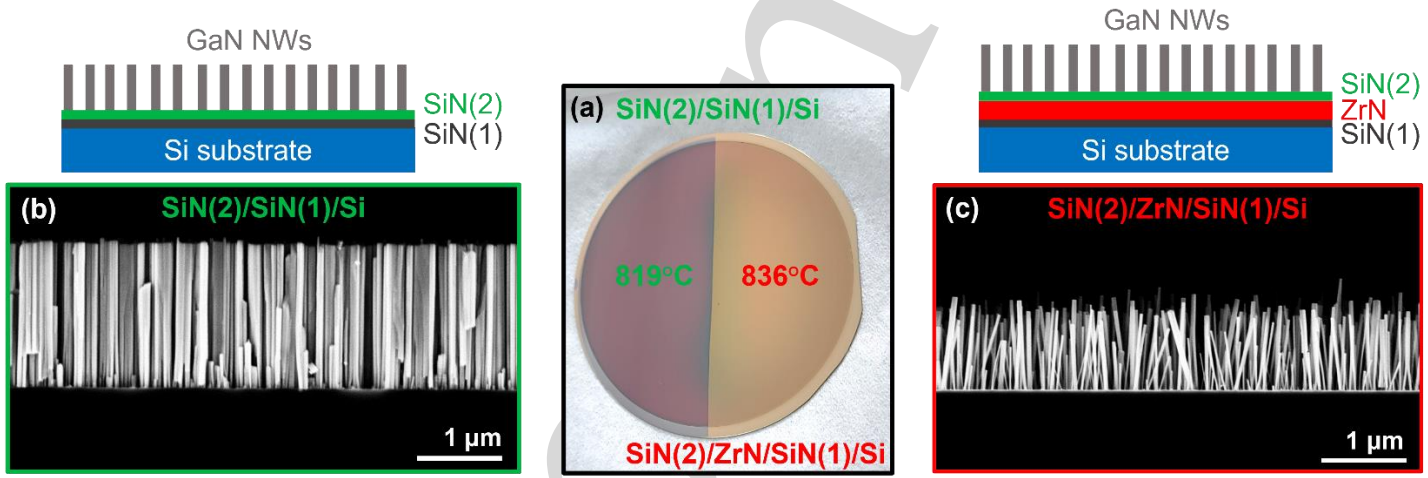

Figure 1. (a) Post growth photo of 3" Si(111) wafer with one half covered by 40 nm ZrN layer. GaN NWs were grown on the whole substrate area; (b) and (c) Cross-section SEM images of GaN NWs grown on the SiN(2)/SiN(1)/Si half and on the SiN(2)/ZrN/Si half of the substrate, respectively. Schematical drawings of the structures are shown above each SEM image.

To demonstrate the influence of a thin ZrN layer on the substrate temperature and thus on self-induced nucleation, we prepared a dedicated substrate for GaN NW growth. Three inch Si(111) wafer was first annealed in the PAMBE growth chamber to remove surface oxides. Next, the substrate was exposed to nitrogen flux to form a protective SiN layer (denoted later on as SiN(1)) on the surface at ~800°C [37]. The aim was to protect the Si wafer against oxidation during the next processing step. Afterwards, the wafer was removed from the growth chamber and 40 nm thick ZrN layer was sputtered only on its one half using a shadow mask. The wafer was subsequently transferred back into the growth system where ~ 2 nm thick layer of SiN (denoted as SiN(2)) was deposited on the whole substrate at room temperature using Si and N sources installed in the MBE system. The role of this SiN(2) film is to bury surface of the ZrN layer and ensure the same mechanism of GaN nucleation on the entire substrate. Thus, any difference in the dimensions and density of the NWs on both parts of the substrate should directly reflect the influence of ZrN layer on the surface temperature of the substrate for exactly the same heating power. Without the SiN(2) film, the NWs nucleating directly on the ZrN layer would be epitaxially linked to the randomly oriented grains of the polycrystalline layer underneath. As a result they would grow following the geometrical selection principle as we have shown previously [26], whereas NWs growing on a nitridated Si wafer would be perpendicular to the substrate surface [38]. Additionally, the incubation times of the NWs on both substrate parts would differ strongly, making direct comparison of their growth kinetics impossible.

After the substrate was heated to the required temperature, but before GaN NW growth started, the optical pyrometer was shifted away from the substrate center to measure temperature of the ZrN-free part equal to 819°C. On the contrary, the temperature of the ZrN-covered half of the wafer could not be accurately determined by optical pyrometry since emissivity of such substrate remains unknown. Therefore, we analyzed kinetics of NW nucleation which is a sensitive probe of even small changes in the substrate surface temperature. For that, we used reflection high-energy electron diffraction (RHEED) following the protocol described in detail in our previous work [13] to measure GaN NW incubation time $\tau$ on both parts of the substrate. The value $\tau = 12$ min measured on the ZrN-free part of the substrate corresponds well to that predicted for N and Ga fluxes used and the growth temperature of 819°C by the analysis of self-induced nucleation of GaN NWs on nitridated Si [14,29]. The same analysis predicts that the incubation time $\tau = 57$ min measured by RHEED on ZrN-coated part of the substrate should correspond to the growth temperature of 836°C.

A photograph of the sample after the NW growth is shown in Fig. 1a. Two distinct halves can be distinguished, with the characteristic golden color of ZrN clearly visible on the right. Typical cross-section images taken on both parts of the substrate are shown in Figs. 1b and 1c for the SiN(2)/SiN(1)/Si and SiN(2)/ZrN/Si parts, respectively. As seen, the nanowires grown on SiN(2)/SiN(1)/Si are of equal length (~1620 nm) and form a densely packed array, whereas the nanowires that grew on SiN(2)/ZrN/Si are generally much shorter (~950 nm) and of varying length, which indicates longer incubation time and length equilibration stage being far from complete. We note that if the NW growth is limited by N flux, the delay of 45 min in GaN nucleation as measured by RHEED should lead to ~675 nm shorter NWs on the ZrN-coated part of the substrate, which exactly corresponds to the difference in NW lengths that is shown on the SEM images in Figs. 1b and 1c.

Our measurements show that presence of 40 nm thick ZrN film increases the substrate temperature by as much as 17ºC. Since the NW incubation time exponentially increases with the growth temperature [13,14,28,29], so large temperature change significantly affects dimensions and density of the NWs for exactly the same heater power. Therefore, for precise controlling PAMBE growth of NW ensembles it is critically important to elaborate a contactless method to determine the true temperature of custom substrates coated with exotic materials with unknown optical parameters, such as ZrN buffer layers of various thicknesses.

In this work we developed a method to determine the real emissivity of a substrate via melting of aluminum (Al) wires as schematically illustrated in Fig. 2a. First, a few Al wires with a diameter of 75 μm were wire-bonded to the substrate surface (Fig. 2b). The as-prepared substrate was subsequently transferred to the PAMBE growth chamber and illuminated by a defocused electron beam from the RHEED gun. After adjusting the electron beam so that shadows of all the wires were clearly visible on the RHEED screen (Fig. 2c), the substrate was gradually heated at a rate of 1°C/min. Once the wires started to disappear indicating their melting, the emissivity of the pyrometer was adjusted to obtain a pyrometer reading of the Al melting point of 660.3°C. An important advantage of the method is that the emissivity of the substrate is measured at the temperature that is close to the temperature range of ~750°C, at which GaN layers and nanostructures are typically grown by PAMBE. Additionally, knowing the positions of the Al wires and observing their subsequent melting allows estimation of the temperature distribution along the substrate diameter.

The technique presented above is universal and can be used for virtually any substrate, especially for one with unknown emissivity, multilayer materials, etc. The only limitation is that the substrate material cannot react with aluminum at temperatures lower than the melting point of Al. Overall the technique is very easy to use and is compatible with the high sterility of MBE systems. Importantly, the emissivity is calibrated for the same substrate and the light optical path configurations that are utilized in the growth process.

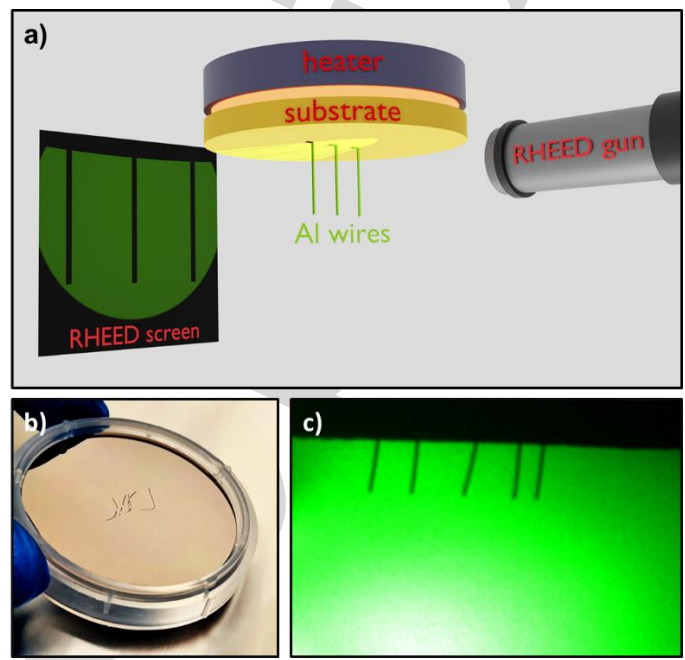

Figure 2. (a) Geometry of the system used for observation of melting of Al wires; (b) Photograph of a substrate wafer with Al wires bonded to its surface; (c) RHEED screen image showing shadows of five separate Al wires attached to the substrate.

The calibration technique presented above was used to determine the emissivity of silicon and sapphire substrates (c-plane $Al_2O_3$ with a backside heat absorbing Mo layer) covered with thin ZrN buffer layers of various thicknesses. The obtained emissivity values as a function of the ZrN buffer thickness are plotted in Fig. 3, with red and blue points for Si and sapphire carrier wafers, respectively. The emissivity of plain Si (red star) was obtained by observing the 7x7 → 1x1 RHEED pattern transition at 860°C [28,39,40]. Our value $\varepsilon(Si) = 0.646$ is very close to that reported in the literature [40]. Fig. 2 shows that the emissivity of both substrates decreases drastically with increasing ZrN layer thickness up to ~100 nm and then saturates. This is in line with theoretical calculations [41], which show that metallic films thicker than 100 nm exhibit bulk emissivity, whereas the emissivity of thinner films strongly increases with decreasing film thickness. In contrast, the same calculations predict that the emissivity of dielectric films increases with increasing layer thickness and saturates for layers thicker than ~10 mm up to macroscopic sizes.

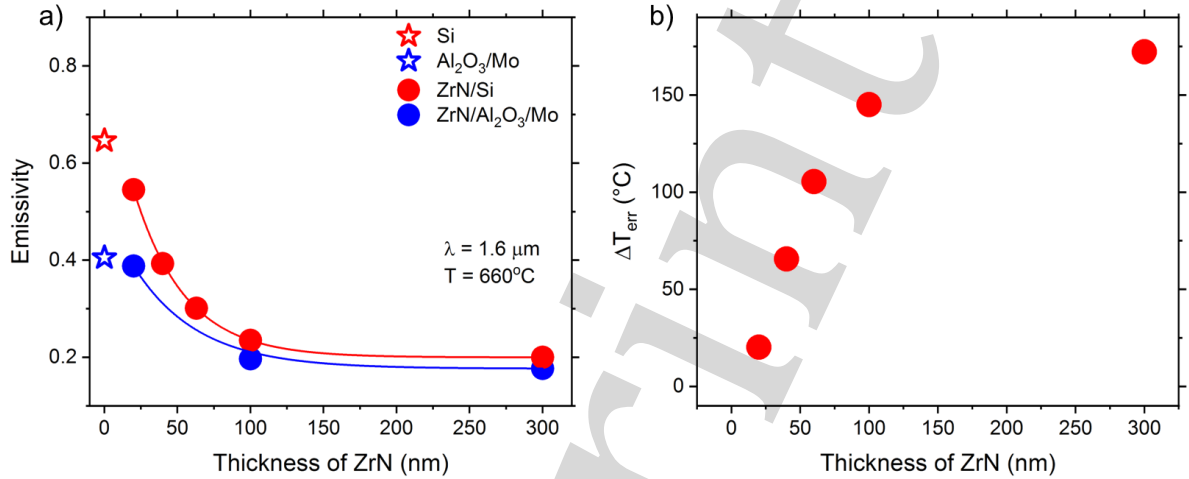

Figure 3. (a) Dependence of substrate emissivity on the ZrN film thickness deposited on silicon (red) and sapphire (blue) substrates measured at T = 660°C and for wavelength λ = 1.6 μm. The lines are drawn to guide the eye. (b) Plot of the $\Delta T_{err}$ parameter versus the thickness of ZrN layer.

The strong emissivity-thickness dependence presented in Fig. 3a has vital implications for the PAMBE growth of GaN nanostructures on thin metallic layers. As we have shown in our previous work [13] under specific growth conditions, even a 10°C change in substrate temperature can result in an increase of the nanowire incubation time from 30 minutes to a few hours. Such temperature measurement errors can be easily caused by a slight misalignment of the substrate emissivity. To put these factors into the perspective of real growth conditions, Fig. 3b shows the error $\Delta T_{err}$ of temperature reading that would be made if for temperature measurements of ZrN/Si substrates the emissivity of bulk Si were used instead of the real emissivity values presented in Fig. 3a. The $\Delta T_{err}$ value is calculated from the equations:

$$\Delta T_{err} = 800°C - T(\varepsilon) \qquad (1)$$

and

$$\varepsilon(Si) \cdot P(\lambda, 800°C) = \varepsilon \cdot P(\lambda, T(\varepsilon)), \qquad (2)$$

where $P(\lambda, T)$ is intensity of black-body radiation described by the Planck's law, while $T(\varepsilon)$ is temperature of ZrN/Si substrate measured by the optical pyrometer for the emissivity $\varepsilon$ that varies with the ZrN layer thickness as shown in Fig. 3a.

Figure 3b presents the values of $\Delta T_{err}$ calculated from eqs. (1) and (2) for the wavelength $\lambda = 1.6$ µm and 800ºC as the reference temperature of the silicon substrate. As shown, the substrate temperature readout error strongly increases for ZrN buffer thickness up to 100 nm, which is the thickness range most interesting for applications as buried mirror or bottom electric contact. Importantly, $\Delta T_{err}$ values are very large, which indicates that using emissivity of the bulk carrier wafer and neglecting the influence of the thin film may lead to huge errors in the substrate temperature measurement and consequently to the loss of control of the growth process. This example clearly shows that proper determination of the emissivity of the utilized substrate is essential for the controlled growth of nanowires. Changing the thickness of the metallic buffer layer must always be followed by a reevaluation of the value of the substrate emissivity. Otherwise large differences in growth temperatures between samples, and consequently a large variation of their properties can be expected for the same power of the substrate heater.

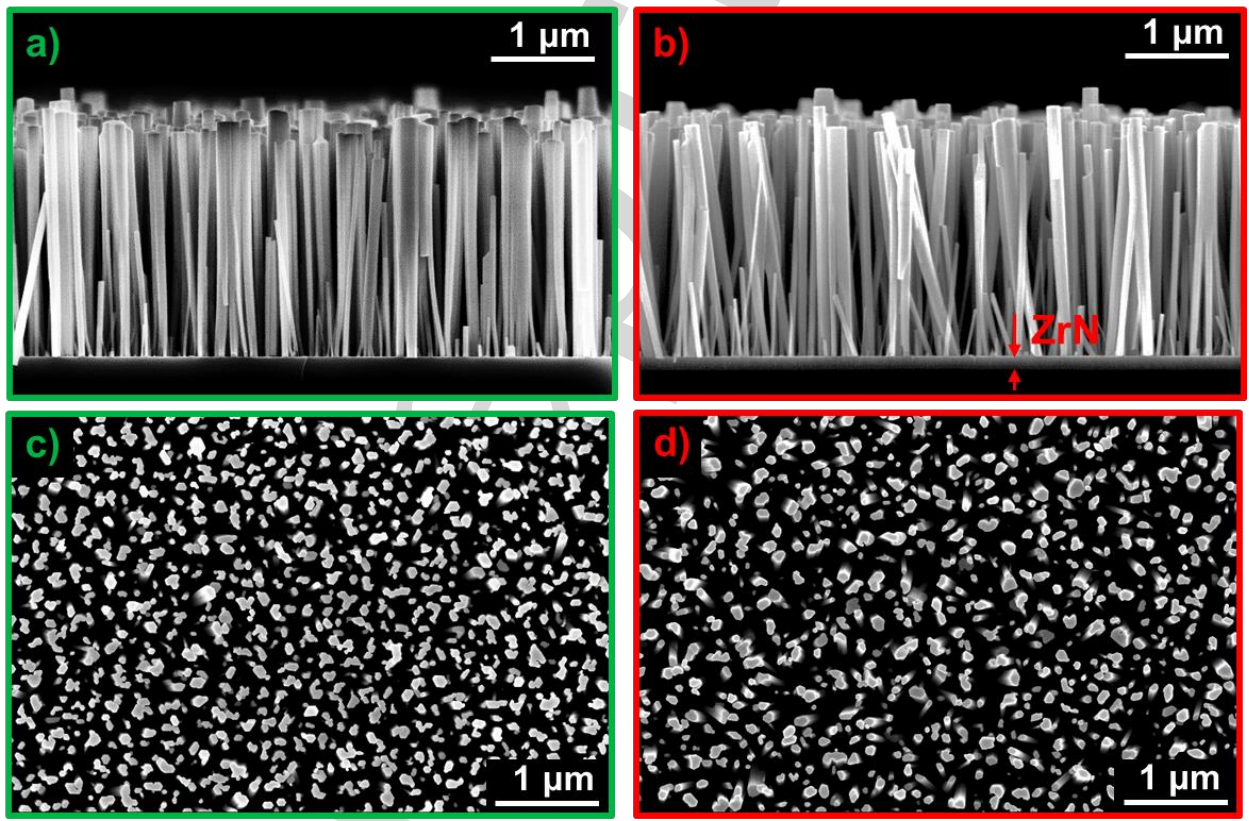

Figure 4. Cross-section (a, b) and top view (c, d) SEM images of two GaN NW arrays that were grown at the same temperature of 838ºC but on different substrates: SiN(2)/SiN(1)/Si (green frame) and SiN(2)/ZrN/Si(1)/Si (red frame).

To demonstrate the accuracy of our temperature calibration technique we grew GaN NWs on two different substrates with the substrate heater power adjusted to obtain exactly the same substrate temperature of 838ºC as measured by optical pyrometry. One was a 3" Si(111) wafer with a thin SiN(2)/SiN(1) layer formed in the same way as described above. The other was also a 3" Si(111) substrate, but in addition a 100 nm ZrN layer was sputtered on its surface before the ~2 nm thick SiN(2) layer was deposited inside the growth chamber. The substrate emissivity for the 100 nm thick ZrN layer on Si was previously calibrated as 0.235 (see Fig. 3). GaN nanowires were grown on these substrates with the same Ga and N fluxes as before and the growth time of 165 min.

Cross-section and top view SEM images of these samples are shown in Fig. 4. Analysis of the images by ImageJ software [42] shows that both arrays exhibit the same nanowire lengths (~1540 nm) and fill factors (~28%). Moreover, the same NW incubation times have been detected by RHEED for GaN nucleation on both substrates. All this indicates that these two NW arrays must have been grown under exactly the same growth conditions as designed. Consequently, this shows that using emissivity values obtained by the Al wire melting calibration technique allows to compensate for the impact of ZrN on the substrate temperature, measure the real surface temperature and regain control of GaN NW nucleation and growth kinetics. Specifically, by following this procedure, we could reproducibly grow NW arrays of exactly the same length and density irrespective of the presence of a ZrN film deposited on the substrate.

Finally, in terms of the MBE growth operational parameters, we note that to achieve the same substrate surface temperature during NW growth of both samples shown in Fig. 4 the heater power must have been adjusted to 52% and 39% for the substrates without and with the ZrN buffer, respectively. This corresponds to ~100ºC difference in the temperature readings by the thermocouple located close to the heater. So large difference again indicates that the presence of a thin metallic buffer significantly affects the heat distribution in the substrate and must be taken into account for precise control of the growth conditions.

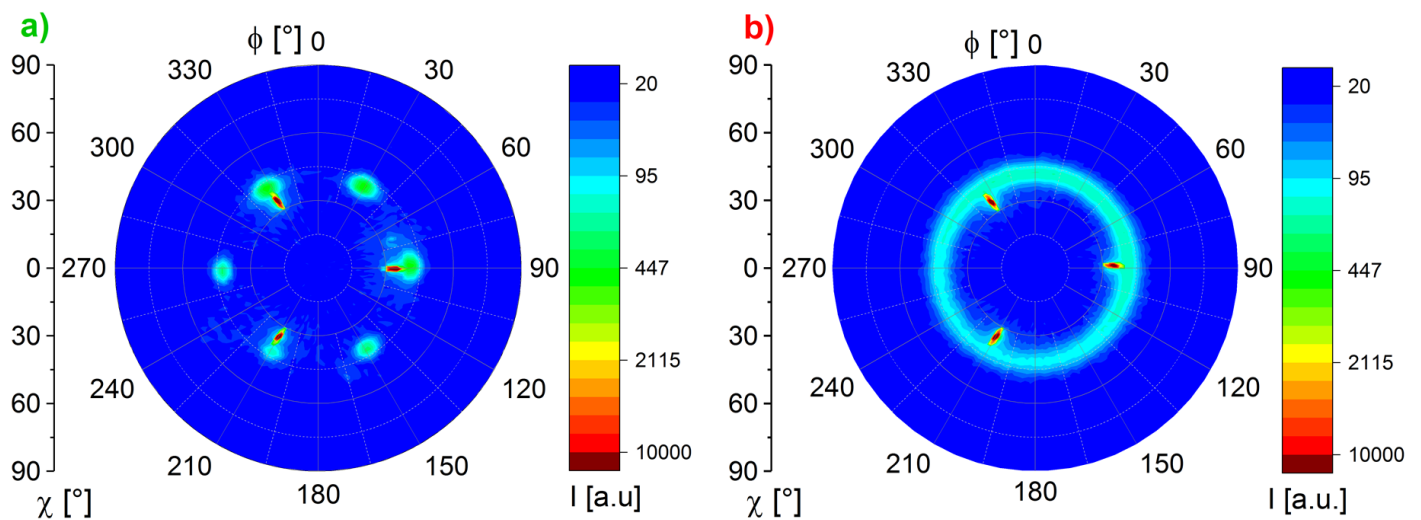

Figure 5. X-ray diffraction 10-12 pole figures of GaN NWs grown on (a) SiN(2)/SiN(1)/Si and (b) SiN(2)/ZrN/SiN(1)/Si substrates at the same growth conditions. Three peak at $\chi = 34°$ originate from Si 220 reflection while those at $\chi = 42°$ come from GaN 10-12 reflection.

The arrangement of GaN NWs on both substrates was studied by X-ray diffraction. The full width at half maximum of the GaN 0002 X-ray rocking curves (not shown) indicates the NW tilt dispersion of ±2° with respect to the substrate normal for the NWs on the SiN(2)/SiN(1)/Si substrate. This value is similar to that commonly reported for GaN NWs on nitridated Si substrates [37]. On the SiN(2)/ZrN/Si substrate, the NW tilt dispersion is a bit larger (±3.2°), probably due to a slightly larger roughness of the ZrN surface than that of Si. Figure 5 shows X-ray diffraction 10-12 pole figures measured for both samples. Three red peaks at $\chi = 34°$ originate from the Si wafer, whereas the signal at $\chi = 42°$ originates from the GaN NWs. As shown in Fig. 5a, the GaN peaks coincide with those from Si, indicating an epitaxial alignment of GaN with the Si substrate despite of the thin SiN layers between them. Importantly, however, this alignment is slightly less pronounced than we previously observed on bare Si(111) substrates nitridated at high temperatures and without the deposition of an additional SiN(2) layer [37]. In contrast, for NWs grown on SiN(2)/ZrN/Si the GaN signal does not depend on the sample rotation angle $\Phi$ forming a circle (see Fig. 5b), which means that the GaN NWs are twisted randomly.

To explain these observations we note that the SiN(1) layer was formed by Si wafer nitridation at high temperature, which should lead to an *in-plane* alignment of nanowires to the Si(111) substrate as discussed in detail in our previous work [37]. Although the SiN(2) film is deposited on top of SiN(1) at low temperature, it apparently partially maintains orientation of the SiN(1) film, which leads to an *in-plane* alignment of nanowires to the substrate, though this alignment is less pronounced than that on SiN(1) alone. In contrast, the SiN(2) film deposited on ZrN at low temperature remains amorphous and the *in-plane* arrangement of the NWs is random, as shown by the circular diffraction pattern in the XRD pole figure in Fig 5b.

**Summary and conclusions**

In this work we study an impact of thin ZrN buffer layer deposited on Si or sapphire wafers on substrate temperature during PAMBE growth. The effect is demonstrated by comparing the self-assembled growth of GaN nanowires on a dedicated Si substrate partly covered by 40 nm thick ZrN buffer. Our analysis shows that the local presence of the ZrN film increases the substrate temperature by as much as 17°C, which significantly affects the NW incubation time and thus the dimensions and density of the NWs for exactly the same heater power.

To quantify the effect we used an optical pyrometry-based temperature calibration technique via melting of aluminum wires attached to the substrate that allows reliable determination of emissivity, and thus precise measurement of surface temperature of custom substrates coated with exotic materials with unknown optical parameters, such as ZrN buffer layers of various thicknesses. Our measurements show that emissivity of ZrN-coated Si and sapphire wafers differs significantly from the bulk ZrN and increases drastically for films thinner than ~100 nm, which agrees with theoretical predictions. The importance of this finding from the perspective of real growth conditions is illustrated by calculating the error $\Delta T_{err}$ of temperature reading that would be made if for temperature measurements of ZrN/Si substrates the emissivity of bulk Si were used instead of the real emissivity values. The $\Delta T_{err}$ values up to ~170°C are obtained, which indicates that using emissivity of the bulk carrier wafer and neglecting the influence of the thin film may lead to huge errors in the substrate temperature measurement and consequently to the loss of control of the growth process.

Then, we demonstrate that by using our temperature calibration procedure we can compensate for the impact of ZrN buffer on substrate temperature and grow identical GaN nanowire arrays on two Si substrates with and without ZrN layers. This finding provides direct evidence that using emissivity values we have measured the desired temperature in separate PAMBE growth processes can be accurately designated on such different substrates and confirms again the usefulness of our approach.

Finally, having two identical arrays of GaN NWs we compare arrangements of the NWs grown on Si and ZrN/Si substrates with a thin SiN nucleation layer. XRD pole figures indicate that on nitridated Si the NWs are *in-plane* aligned to the Si crystal lattice. On the contrary, random NW twist is observed on ZrN-coated Si substrates. This finding points to a partial recrystallization of SiN layer on nitridated Si while it remains amorphous on the ZrN surface.

**Acknowledgements**

This research was partly funded by the Polish National Science Centre NCN grants 2021/43/D/ST7/01936 and 2022/04/Y/ST7/00043 (Weave-Unisono). The authors are grateful to T. Wojciechowski for taking SEM images of the samples.

**Author Contributions**

KO - investigation, data analysis, writing - original draft preparation, ZRZ - writing - review and editing, supervision, resources, project administration, AW - investigation, MG - investigation, MS – investigation, writing - review and editing, methodology, resources, supervision.

**Competing interests**

The authors declare no competing interests.

**Data availability**

Data supporting this article have been uploaded to the data repository: https://doi.org/10.18150/HYOSCB